\documentclass[aps,prd,12pt,preprint,tightenlines,superscriptaddress,
amsfonts,amssymb,amsmath,byrevtex,showpacs,nofootinbib]{revtex4-1}
\usepackage{graphics}
\usepackage{epsfig,subfigure}
\begin{document}
\newcommand{\dR}{\mathbb R}
\newcommand{\dC}{\mathbb C}
\newcommand{\dS}{\mathbb S}
\newcommand{\dZ}{\mathbb Z}
\newcommand{\id}{\mathbb I}
\newcommand{\dM}{\mathbb M}
\newcommand{\dH}{\mathbb H}
\newcommand{\tm}{\tilde{\mu}}
\newcommand{\tn}{\tilde{\nu}}

\title{Quantization of FRW universe via gauge-fixed action}

\author{Przemys{\l}aw Ma{\l}kiewicz
\\ Theoretical Physics Department, Institute for Nuclear Studies
\\ Ho\.{z}a 69, 00-681 Warsaw, Poland;
\\ pmalk@fuw.edu.pl}

\date{\today}

\begin{abstract}
This paper is devoted to investigation of the quantum
Friedman-Robertson-Walker universe with matter satisfying the
equation of state $p=w\rho$, where $w$ is an almost arbitrary
constant. The procedure starts with a reduced Lagrangian, which
describes the system in a gauge fixed, so that the evolution
parameter corresponds to the cosmological time. Then we construct
the phase space, which is believed to correspond to the reduced
phase space consisting of Dirac's observables. The physically
relevant quantities are mapped into operators. We show that the
operators have self-adjoint realizations and that there exist
quantum states for which the evolution across singularity is
well-defined.
\end{abstract}

\pacs{98.80.Qc,04.60.Pp,04.20.Jb}\maketitle

\section{Introduction}

For a gauge system like general relativity, there exist two ways
of quantizing. (i) The Dirac method \cite{PAM} starts with
quantization of all the degrees of freedom of the configuration
space of e.g. the Hilbert-Einstein action. Then the gauge freedom
is removed from the quantum theory by solving the constraint
operator equation. On the other hand, (ii) the reduced phase space
method (see e.g. \cite{Mal}) begins with canonical formulation of
the classical theory and then the gauge freedom is removed by
identification of all Dirac's observables through the following
weak equality: $\{C,\Omega\}\approx 0$, where $C$ is a constraint,
$\Omega$ is a Dirac's observable and the equality is weak in the
sense that it holds on the constraint surface $C=0$ (it excludes
the solutions of the type $\Omega=C$). Then the Dirac observables,
which are physical degrees of freedom, are mapped into quantum
operators.

In this paper we take yet another route: we remove the gauge
freedom already at the level of Lagrangian. Then we move to a
Hamiltonian formulation and subsequently we quantize the canonical
system using the Schr\"odinger representation. The analysis is
restricted to the compact, flat FRW universe.

In section \ref{ct}, starting with a given Lagrangian, we show
that it describes the system under consideration and arrive at
canonical formulation in convenient basic variables. In section
\ref{sq}, we map the phase space functions into operators and
study their self-adjointness. Next, in section \ref{eoevopc} we
compute the evolution of mean values of the relevant operators. We
conclude in section \ref{d}.

In appendix \ref{rp} we make conjecture that the reduced phase
space method is closely related to the reduced Lagrangian method.

In appendix \ref{B} we report on an unsuccessful attempt at
resolving the singularity. It is discussed and compared with
approach used throughout this paper in concluding section \ref{d}.

\section{Classical theory}
\label{ct} The metric of the flat FRW universe in the comoving
coordinates and with respect to the cosmological time $t$ reads:
\begin{equation}\label{FRW}
    ds^2=-dt^2+a(t)^2d\overline{x}^2,
\end{equation}
where $a(t)$ is a scale factor and $d\overline{x}$ is a distance
measure in a space-like leaf $\Sigma$. Instead of the scale factor
$a$, we are going to consider the dynamics of the physical length
between two unspecified points in $\Sigma$ and we denote it by
$l(t):=l_0\frac{a(t)}{a(t_0)}$, where $l_0$ is the length at the
moment $t=t_0$. Therefore, for each value of $t$, the length
$l(t)$ is a physical quantity, i.e. the Dirac observable. Now let
us see how one can introduce the dynamics of $l$.

\subsection{Lagrangian formulation}
Let us examine the following action integral:
\begin{equation}\label{action}
    S=\int\mathbb{L}(l,\dot{l})~dt=\alpha\int
    l^{3w+1}\dot{l}^2~dt,
\end{equation}
where $\alpha:=\frac{3}{8\pi G}\cdot\frac{1}{\lambda^{3w}}$ and
$\lambda$ is a constant of the length dimension and will be
specified later. The parameter $w$ is a real constant. Variation
with respect to $l$ gives the equation of motion:
\begin{equation}
-(3w+1)\alpha l^{3w}\dot{l}^2-2\alpha
l^{3w+1}\ddot{l}=0=\bigg(l^{\frac{3w+3}{2}}\bigg),_{tt},
\end{equation}
which has the solutions:
\begin{equation}\label{sol}
l(t)=l_0\bigg(\frac{3w+3}{2}\frac{\dot{l}_0}{l_0}(t-t_0)+1\bigg)^{\frac{2}{3w+3}},
\end{equation}
satisfying the initial conditions:
\begin{equation}\label{int}
l(t_0)=l_0,~~~~\dot{l}(t_0)=\dot{l}_0.\end{equation} Let us turn
to the Friedman equations:
\begin{eqnarray}
H^2&=&\frac{8\pi G}{3}\rho,\\
H^2+\dot{H}&=&-\frac{4\pi G}{3}(\rho+3p),
\end{eqnarray}
where $H$ is Hubble's parameter, $\rho$ is the energy density and
$p$ is the pressure of the matter in the universe. Now, making use
of $H:=\frac{\dot{l}}{l}$, we find out that:
\begin{equation}
    H^2=(\dot{l}_0^2l_0^{3w+1})\frac{1}{l^{3w+3}},~~~~H^2+\dot{H}=-\frac{1}{2}(1+3w)(\dot{l}_0^2l_0^{3w+1})\frac{1}{l^{3w+3}},
\end{equation}
and we easily conclude that our system given by the action
integral (\ref{action}) is the flat FRW universe with $\frac{8\pi
G}{3}\rho=(\dot{l}_0^2l_0^{3w+1})\frac{1}{l^{3w+3}}$ and
$p=w\rho$. \footnote{Note that these equations refer to the
specific choice of evolution parameter, which is the cosmological
time, and hence there are no gauge degrees of freedom in this
formulation.}

Hence, we have succeeded in the gauge-fixed Lagrangian formulation
of the dynamics of the flat FRW universe filled with matter
satisfying the equation of state $p=w\rho$. The connection of this
reduced Lagrangian with the reduced phase space procedure, which
begins with the introduction of the kinematical phase space
(consisting of both gauge and physical degrees of freedom), is
commented in appendix \ref{rp}.

The energy of matter contained in a fiducial cubic cell of the
edge length changing with time as $l(t)$, reads:
\begin{equation}
 E_{matt}=\frac{3l_0^3}{8\pi
G}\bigg(\frac{\dot{l}_0}{l_0}\bigg)^2\bigg(\frac{l}{l_0}\bigg)^{-3w},
\end{equation}
and, for $w\neq 0$, is clearly not conserved. For instance, the
matter may consist of particles, which move with respect to the
comoving reference frame, then the particles loose/gain their
kinetic energy as the spacetime expands/shrinks and the energy of
matter changes with the evolution of the universe. However, one
may introduce a conserved quantity, the total energy:
\begin{equation}\label{energy}
E_{tot}=\frac{1}{\lambda^{3w}}\rho\cdot l^{3w+3}, \end{equation}
which is equal to the energy of matter at the moment when the
value of the length of the fiducial cubic cell is equal to
$\lambda$. If the moment happens to be $t=t_0$, then
$E_{tot}=E_{matt}(t_0)=\frac{3l_0^3}{8\pi
G}\big(\frac{\dot{l}_0}{l_0}\big)^2$, and $\lambda=l_0$.

\subsection{Canonical formulation}
We perform the Legendre transformation:
\begin{equation}
l\mapsto l\in\mathbb{R}_+,~~~~\dot{l}\mapsto
p_l:=\frac{\mathrm{d}\mathbb{L}}{\mathrm{d}\dot{l}}=2\alpha
l^{3w+1}\dot{l}\in\mathbb{R},
\end{equation}
so that
\begin{equation}
\mathbb{H}=p_l\dot{l}-\mathbb{L}=\frac{p_l^2}{4\alpha l^{3w+1}}~~~~\textrm{and}~~~~\{l,p_l\}=1
\end{equation}
Hamilton's equations:
\begin{equation}\label{Heq}
\dot{l}=\{l,\mathbb{H}\}=\frac{p_l}{2\alpha
l^{3w+1}},~~~~\dot{p}_l=\{p_l,\mathbb{H}\}=(3w+1)\frac{p_l^2}{4\alpha
l^{3w+2}}.
\end{equation}
can be integrated and we obtain:
\begin{equation}\label{sol1}
p_l=Cl^{\frac{3w+1}{2}},~~\textrm{or}~~p_l=0,~l=C~\textrm{(stationary
universe)},
\end{equation}
where $C$ is any real constant. The solutions split into three
cases: $w<-1/3$, $w=-1/3$ and $w>-1/3$. They respectively
correspond to the momentum of space blowing up, being constant and
vanishing as the universe approaches the singularity $l=0$.

We are going to study physically interesting quantities like
Hubble's parameter or energy density, which take the following
form, respectively:
\begin{equation}
H=\frac{p_l}{2\alpha l^{3w+2}},~~~~\rho=\frac{3}{8\pi G}H^2.
\end{equation}
The Hamiltonian $\mathbb{H}$ is a conserved quantity like the
total energy of the system defined in eq. (\ref{energy}). The
comparison of these two leads to the condition:
\begin{equation}
 \frac{1}{\alpha}\frac{p_l^2}{4 l^{3w+1}}=\mathbb{H}=E_{tot}=\frac{1}{\alpha^2}\frac{3}{8\pi G}\frac{p_l^2}{4
 l^{3w+1}}\frac{1}{\lambda^{3w}},
\end{equation}
and hence
\begin{equation}
 \alpha=\frac{3}{8\pi G\lambda^{3w}},
\end{equation}
which is in accordance with our prior definition and explains the
physical meaning of $\lambda$.

\subsection{Canonical transformation} In what follows we will look for a more convenient conjugate pair.
We assume that:
\begin{enumerate}
    \item the canonical transformation is valid globally
    \item the 'stationary universes' in (\ref{sol1}) are included in the system and thus well-mapped by the
    transformation
    \item canonical transformation does not extend the phase
    space, in particular it preserves the topology of the phase
    space under consideration
\end{enumerate}
Now, let us break the above rules in the following example:
\begin{equation}
\{l,p_l\}\longrightarrow\{\tilde{l}:=\mathbb{H},\tilde{p}_l:=\frac{2H^{-1}}{3(w+1)}\}~,
~~w\neq -1.
\end{equation}
The map is singular for $p_l=0$ and thus the 'stationary
universes' being the points at the axis $p_l=0$, are excluded now.
Instead, the closure of the range of the map covers a new region
of the space of parameters, $\tilde{p}_l=0$. As a consequence, the
trajectories in the phase space are complete and the singularities
are absent! But this is not a good way to resolve the singularity,
since, apart from the lack of quantum physics, the procedure has a
drawback in that it is non-unique and here are further examples:
\begin{equation}
\{l,p_l\}\longrightarrow\{\tilde{l}:=\mathbb{H},\tilde{p}_l:=\frac{2H^{-1}}{3(w+1)}+\textrm{sgn}(p_l)f(\mathbb{H})\}~,
~~w\neq -1,~~f~\textrm{arbitrary}.
\end{equation}
Therefore, making use of a `regular' map, we introduce the
following new canonical pair:
\begin{equation}
\{l,p_l\}\longrightarrow\{\tilde{l}:=\frac{4}{3w+3}l^{(3w+3)/2},\tilde{p}_l:=\frac{p_l}{2l^{(3w+1)/2}}\}~,
~~w\neq -1.
\end{equation}
Now, the Hamiltonian reads:
\begin{equation}
\mathbb{H}=\frac{1}{\alpha}\tilde{p}_l^2,
\end{equation}
the Hubble parameter:
\begin{equation}
H=\frac{1}{\alpha}\frac{4}{3w+3}\frac{\tilde{p}_l}{\tilde{l}},
\end{equation}
the energy density:
\begin{equation}
\rho=\frac{1}{\alpha^2}\frac{2}{\pi G
(3w+1)}\frac{\tilde{p}_l^2}{\tilde{l}^2},
\end{equation}
and the length:
\begin{equation}
l=\big({\frac{3w+3}{4}~\tilde{l}}\big)^{2/(3w+3)}.
\end{equation}
\section{Schr\"odinger quantization}
\label{sq}

In the Schr\"odinger representation:
\begin{equation}
\tilde{l}\mapsto\hat{l}:=x,~~~~\tilde{p}_l\mapsto\hat{p}_l:=-i\hbar\frac{d}{dx},
\end{equation}
and we get the following formally self-adjoint operators that are
relevant in the cosmological context and will be studied below:
\begin{eqnarray}
&&
\mathbb{H}\mapsto\hat{\mathbb{H}}=-\frac{\hbar^2}{\alpha}\frac{d^2}{dx^2},\\
\label{h}
&& H\mapsto\hat{H}=\frac{\hbar}{\alpha}\frac{4}{3w+3}(-i\frac{d}{dx}x^{-1}-ix^{-1}\frac{d}{dx}),\\
&& \rho\mapsto\hat{\rho}=\frac{\hbar^2}{\alpha^2}\frac{1}{2\pi G
(3w+1)}(-i\frac{d}{dx}x^{-1}-ix^{-1}\frac{d}{dx})^2,\\
&&  l\mapsto\hat{l}=\big({\frac{3w+3}{4}x}\big)^{2/(3w+3)}.
\end{eqnarray}
The range of $x$ is $(0,\infty)$. Thus, the problem of evolution
across singularity is formulated in terms of a freely moving
quantum particle on a half-line.

\subsection{Hamiltonian and the evolution}

The theory of the operator $-\frac{d^2}{dx^2}$ on half-line may be
found in \cite{Der}. In short, there are infinitely many ways, in
which one may define the operator $\hat{\mathbb{H}}$ to be
essentially self-adjoint on $L^2(R_+,dx)$ and we will denote these
options by $\hat{\mathbb{H}}_{\mu}$ for $\mu\in
\mathbb{R}\cup\{\infty\}$\footnote{Usually, the choices $\mu=0$
and $\mu=\infty$ correspond to the Neumann and Dirichlet
conditions, respectively.}. In order to calculate the evolution of
a given state one may apply the appropriate Fourier Transform,
denoted by $FT^{\mu}$, and multiply the resultant wave-function by
$e^{-ik^2t}$, where $k$ is the momentum.

The important point to be made here is that we {\bf do} introduce
an evolution operator. In the quantization scheme proposed in
\cite{Mal}, one only quantizes the Dirac observables enumerated by
a classical evolution parameter - one does not introduce a
generator of change in this parameter, called the true
Hamiltonian, in order to quantize it. However, in our model of
universe this approach would not work - it would not give any
sensible solution to the singularity problem. The crucial
difference is that in the model presented in \cite{Mal} the
classical dynamics was already non-singular, whereas in the model
considered here the dynamics ends at the point $l=0$.

The quantization of the evolution may give some hopes to resolve
the singularity. The following heuristic argument shows that the
closer to the singularity the more the quantum universe departures
from the classical one:
\begin{eqnarray}\label{cl}
  \{H,\mathbb{H}\} &\sim& -2H^2, \\ \label{qu}
  \frac{1}{i}[\hat{H},\hat{\mathbb{H}}] &\sim&
  -2\hat{H}^2+\frac{1}{2}x^{-4}.
\end{eqnarray}
We observe that at the classical level (\ref{cl}) the value of the
Hubble parameter $H$ goes to minus infinity as the universe
approaches the singularity, whereas at the quantum level
(\ref{qu}) the term $x^{-4}$ opposes the unbounded decrease of the
operator $\hat{H}$. From the form of $\hat{H}$ given in (\ref{h}),
one sees that the `closer' the wave-function to $x=0$, the bigger
the value of $\hat{H}^2$. But at the same time, the closer the
wave-function to $x=0$, the bigger the value of the opposing term
$x^{-4}$.


\subsection{The energy density and Hubble operators}

We will study the formally self-adjoint Hubble operator:
\begin{equation}\label{hubble}
    \hat{H}=\frac{4}{3w+3}\frac{\hbar}{\alpha}(-i\frac{d}{dx}x^{-1}-ix^{-1}\frac{d}{dx}).
\end{equation}
Let us consider the following inverse mapping
$L^2(\mathbb{R}_+,dx)\mapsto L^2(\mathbb{R}_+,dy)$:
\begin{equation}\label{isomap}
    \psi(x)\mapsto \phi (y):=
    \frac{\psi(2\sqrt{y})}{y^{1/4}},~~y>0.
\end{equation}
Let us see that this mapping is isometric, and hence unitary:
\begin{equation}\label{isomapPROOF}
    \int_{0}^{\infty}\overline{\psi}(x)\psi(x)~dx=\int_{0}^{\infty}\overline{\psi}(2\sqrt{y})\psi(2\sqrt{y})~2d\sqrt{y}=\int_{0}^{\infty}\frac{\overline{\psi}(2\sqrt{y})}{y^{1/4}}
    \frac{\psi(2\sqrt{y})}{y^{1/4}}~dy.
\end{equation}
Application of the mapping (\ref{isomap}) to the operator
$\hat{H}$ gives:
\begin{equation}\label{opmap}
\frac{-4i}{3w+3}\frac{\hbar}{\alpha}y^{-1/4}(-x^{-2}+2x^{-1}\frac{d}{dx})y^{1/4}=\frac{-4i}{3w+3}\frac{\hbar}{\alpha}\frac{d}{dy},
\end{equation}
which is a simple derivative operator. Analogically, under the
same mapping the energy density operator $\hat{\rho}$ is
transformed into Laplacian $\sim -\frac{d^2}{dy^2}$.

Again, there are infinitely many inequivalent ways, in which one
may define a self-adjoint $\hat{\rho}$, and we will denote these
options by $\hat{\rho}_{\mu},~\mu\in \mathbb{R}\cup\{\infty\}$.
However, the Hubble operator, being a momentum operator, cannot be
self-adjoint on the half-line. Therefore, we will redefine
$\hat{H}:=\sqrt{\hat{\rho}}$. The spectrum of these operators
reads $sp(\hat{\rho}_{\mu})=sp(\hat{H}_{\mu})=\mathbb{R}_+$ for
any $\mu$.
\subsection{The volume operator}

The volume operator $\hat{v}:=\hat{l}^3=\big({\frac{3w+3}{4}x}\big)^{2/(w+1)}$ is a self-adjoint operator on $L^2(\mathbb{R}_+,dx)$ and its
spectrum for $w>-1$ reads $sp(\hat{v})=\mathbb{R}_+$.

\section{Evolution of physical quantities}
\label{eoevopc}

Let us restrict to the study of the evolution that preserves the
Dirichlet condition (i.e. generated by $\hat{\mathbb{H}}_{\mu}$
for $\mu=\infty$). We pick the following family of wave-functions:
\begin{equation}
\tilde{\phi}_a(k):= Nke^{-ak^2},~~k>0,
\end{equation}
where $N=2\frac{(2a)^{3/4}}{\pi^{1/4}}$, $[k]=[x]^{-1}=m^{-\frac{3w+3}{2}}$, $[a]=m^{3w+3}$ and evolve them in time:
\begin{equation}
\exp{(-\frac{i}{\hbar}t\hat{\mathbb{H}})}\tilde{\phi}_a(k)=\exp{(-i\frac{\hbar}{\alpha}tk^2)}\tilde{\phi}_a(k)=Nke^{-(a+i\frac{\hbar}{\alpha}t)k^2}=\tilde{\phi}_a(k,t).
\end{equation}
Applying the Fourier sine transform $FT^{\infty}$ renders:
\begin{equation}\label{st}
 \phi_a(x,t):=\int_0^{\infty}\sin(kx)\tilde{\phi}_a(k,t)dk=\frac{Nx}{(2(a+i\frac{\hbar}{\alpha}t))^{3/2}}~e^{-\frac{x^2}{4(a+i\frac{\hbar}{\alpha}t)}}.
\end{equation}
\subsection{The invariant $\sqrt[w+1]{E_{tot}}$}

Let us calculate the expectation value of the operator
corresponding to the classical invariant $\sqrt[w+1]{E_{tot}}$~:
\begin{eqnarray}\nonumber
    \langle \sqrt[w+1]{E_{tot}}~\rangle:=\langle\tilde{\phi}_a(k)|\sqrt[w+1]{\hat{\mathbb{H}}}~\tilde{\phi}_a(k)\rangle=\sqrt[w+1]{\frac{\hbar^2}{\alpha}}|N|^2
    \int_0^{\infty}e^{-2ak^2}k^{\frac{2w+4}{w+1}}~dk\\ \label{einv}
    =\sqrt[w+1]{\frac{\hbar^2}{2\alpha}\bigg(\frac{2}{\sqrt{\pi}}\Gamma\bigg(\frac{5+3w}{2+2w}\bigg)\bigg)^{w+1}
    \cdot\frac{1}{a}}~.
\end{eqnarray}
It will turn out to be useful for linking the classical
trajectories with quantum states $\phi_a$.

\subsection{Volume}
The expectation value for the volume operator for the state
$\phi_a(x,t)$ equals:
\begin{eqnarray}\nonumber
    &&\langle v\rangle(t)=\frac{N^2}{8(a^2+\frac{\hbar^2}{\alpha^2}t^2)^{3/2}}\bigg({\frac{3w+3}{4}}\bigg)^{2/(w+1)}\int_0^{\infty}x^{\frac{2(w+2)}{w+1}}e^{-\frac{a
    x^2}{2(a^2+\frac{\hbar^2}{\alpha^2}t^2)}}~dx=\\\label{volume}
    &&\frac{N^2}{8(a^2+\frac{\hbar^2}{\alpha^2}t^2)^{3/2}}\bigg({\frac{3w+3}{4}}\bigg)^{2/(w+1)}2^{\frac{1}{2}
    +\frac{1}{1+w}}\bigg(\frac{a}{a^2+\frac{\hbar^2}{\alpha^2}t^2}\bigg)^{-\frac{3}{2}-\frac{1}{1+w}}\Gamma\bigg(\frac{3}{2}+\frac{1}{1+w}\bigg)\\
    \nonumber
    &&=\frac{2^{\frac{w-2}{w+1}}}{\sqrt{\pi}}\bigg(\frac{3w+3}{\sqrt{a}}\bigg)^{\frac{2}{w+1}}\Gamma\bigg(\frac{5+3w}{2+2w}\bigg)(a^2+\frac{\hbar^2}{\alpha^2}t^2)^{\frac{1}{1+w}}.
\end{eqnarray}
One observes that at $t=0$ the volume acquires its minimal value
\begin{equation}\label{volcrt}
v_{crit}:=a^{\frac{1}{1+w}}\frac{2^{\frac{w-2}{w+1}}}{\sqrt{\pi}}(3w+3)^{\frac{2}{w+1}}\Gamma\bigg(\frac{5+3w}{2+2w}\bigg).
\end{equation}
Let us study the asymptotic behavior in (\ref{volume}) for
$t\rightarrow \pm\infty$:
\begin{equation}\label{qv}
    \langle v\rangle(t)\simeq\frac{2^{\frac{w-2}{w+1}}}{\sqrt{\pi}}(3w+3)^{\frac{2}{w+1}}\Gamma\bigg(\frac{5+3w}{2+2w}\bigg)\bigg(\frac{\frac{\hbar^2}{\alpha^2}t^2}{a}\bigg)^{\frac{1}{w+1}},~~\frac{\hbar^2}{\alpha^2}t^2\gg
    a^2.
\end{equation}
On the other hand, the classical solution in (\ref{sol}) tells
that the volume of the fiducial cubic cell with the edge $l$
evolves as:
\begin{equation}\label{vol}
    v(t)=l^3_0\bigg(\frac{3w+3}{2}(t-t_0)\frac{\dot{l}_0}{l_0}+1\bigg)^{\frac{2}{w+1}},
\end{equation}
which for $t-t_0\gg\frac{2}{3w+3}\frac{l_0}{\dot{l}_0}$ reads:
\begin{equation}\label{cv}
    v(t)\simeq l^3_0\bigg(\frac{\dot{l}_0}{2l_0}\bigg)^{\frac{2}{w+1}}(3w+3)^{\frac{2}{w+1}}(t^2)^{\frac{1}{w+1}}
    =\sqrt[w+1]{\frac{E_{tot}}{\alpha}}\bigg(\frac{3w+3}{2}\bigg)^{\frac{2}{w+1}}t^{\frac{2}{w+1}}.
\end{equation}
The comparison of (\ref{cv}) with (\ref{qv}) gives the following relation:
\begin{equation}\label{qini}
    a=\frac{\hbar^2}{2\alpha}\bigg(\frac{2}{\sqrt{\pi}}\Gamma\bigg(\frac{5+3w}{2+2w}\bigg)\bigg)^{w+1}\frac{1}{E_{tot}},
\end{equation}
which is in accordance with (\ref{einv}), if we substitute
$E_{tot}$ for $\langle\sqrt[w+1]{E_{tot}}\rangle^{w+1}$. Plugging
this relation into the formula (\ref{volcrt}) we find out that at
the bounce the critical volume of the cubic cell reads:
\begin{equation}\label{v}
v_{crit}=\frac{4}{\pi}\Gamma^2\bigg(\frac{5+3w}{2+2w}\bigg)\bigg(\frac{3w+3}{4}\bigg)^{\frac{2}{w+1}}\sqrt[w+1]{\frac{\hbar^2}{\alpha
E_{tot}}}~.
\end{equation}
First, notice that $v_{crit}$ depends only on the classical
invariant of motion $E_{tot}$, so it does not depend on the
particular moment of evolution of the classical universe, at which
we calculate $v_{crit}$. Next, notice that it is proportional to
$\frac{1}{\sqrt[w+1]{E_{tot}}}$, hence \emph{the more massive the
universe the smaller its critical size}\footnote{The more massive
universe means the universe of bigger size with the same energy
density of matter or the universe of the same size with higher
energy density of matter.}. The formula (\ref{v}) knows nothing
about Planck scale and combined with the knowledge of the early
Universe evolution, can lead to the bounds on the mass of the
whole Universe.

\begin{figure}[t]
\centering
\includegraphics[totalheight=2in]{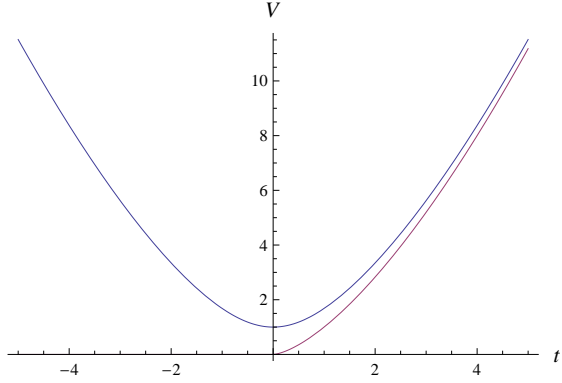}
\caption{The comparison of classical and quantum behavior of the
volume for $w=1/3$, $\alpha=\hbar=a=1$.} \label{fig1}
\end{figure}
\subsection{Energy density and Hubble parameter}
First we make use of the unitary mapping $L^2(\mathbb{R}_+,dx)\mapsto L^2(\mathbb{R}_+,dy)$ and map the state $\phi_a(x,t)$ accordingly to
(\ref{isomap}). We find:
\begin{equation}
    \phi_a(x,t)\mapsto
    \psi_a(y,t):=y^{-1/4}\phi_a(2\sqrt{y},t)=\frac{Ny^{1/4}}{(2(a+it))^{3/2}}~e^{-\frac{y}{(a+it)}}.
\end{equation}
We assume that the action of $\hat{H}$ and $\hat{\rho}$ on
$\psi(y)\in L^2(\mathbb{R}_+,dy)$ preserves the Dirichlet
condition (i.e. we set $\hat{\rho}=\hat{\rho}_{\infty}$). We
perform the Fourier sine transform $FT^{\infty}$ on $\psi_a(y,t)$:
\begin{equation}
\tilde{\psi}_a(k,t)=\int_0^{\infty}\sin(kx)\psi_a(x,t)dk=\tilde{N}\frac{\sin(\frac{5}{4}\arctan((a+it)k))}{(\frac{1}{(a+it)^2}+k^2)^{5/8}},
\end{equation}
where $\tilde{N}=\frac{N\sqrt{\pi}\Gamma(\frac{5}{4})}{2(a+it)^{3/2}}$ and hence we have
\begin{equation}\label{psi2}
    \overline{\tilde{\psi}}_a\tilde{\psi}_a=|\tilde{N}|^2\bigg|\frac{\sin(\frac{5}{4}\arctan(\frac{k}{C}))}{(C^2+k^2)^{\frac{5}{8}}}\bigg|^2,
\end{equation}
where $C=\frac{1}{(a+it)}$. Now, in the momentum representation,
the operators $\hat{H}$ and $\hat{\rho}$ are proportional to $k$
and $k^2$, respectively. Since the function (\ref{psi2}) for large
$k$ behaves like \begin{equation}
\overline{\tilde{\psi}}_a\tilde{\psi}_a\sim
k^{-5/2},\end{equation} only the operators $\hat{O}$ such that
$\hat{O}\leq C k^{\alpha}$ for large $k$, a constant $C$ and
$\alpha<\frac{3}{4}$, are well defined on the states that we
consider in the sense that:
\begin{equation}\langle\hat{O}{\tilde{\psi}}_a|\hat{O}\tilde{\psi}_a\rangle<\infty .\end{equation}
However, we can still consider the expectation values
\begin{equation}\label{expval}
\langle\tilde{\psi}_a|\hat{O}\tilde{\psi}_a\rangle:=\langle
\hat{O}^{\frac{1}{2}}\tilde{\psi}_a|\hat{O}^{\frac{1}{2}}\tilde{\psi}_a\rangle
,
\end{equation}
for operators $\hat{O}\leq C k^{2\alpha}$. The reason is that the
integral $\langle\tilde{\psi}_a|k^{2\alpha}\tilde{\psi}_a\rangle$
is convergent and since the states $\tilde{\psi}_a$ may be
approximated by states from the domain of the operator
$k^{2\alpha}$ up to arbitrary accuracy, the integral
(\ref{expval}) keeps the physical meaning of the expectation value
of this operator.

In particular, we may obtain the integral formula for the expectation value of the Hubble parameter (but not the energy density):
\begin{equation}
\langle\tilde{\psi}_a|k\tilde{\psi}_a\rangle=|\tilde{N}|^2\int_{0}^{\infty}
\bigg|\frac{\sin(\frac{5}{4}\arctan(\frac{k}{C}))}{(C^2+k^2)^{\frac{5}{8}}}\bigg|^2k~dk<\infty
.
\end{equation}
We have already learnt from the study of the volume that the
bounce occurs for the states $\tilde{\psi}_a$ at $t=0$, hence we
may calculate the expectation value of the operator
$\hat{H}=\frac{4}{3w+3}\frac{\hbar}{\alpha}k$ at the bounce:
\begin{equation}\label{Hcrit}
   H_{crit}=\frac{\hbar}{\alpha}\cdot\frac{4|\tilde{N}|^2}{3w+3}a^{\frac{1}{2}}\int_{0}^{\infty}
\bigg|\frac{\sin(\frac{5}{4}\arctan(k))}{(1+k^2)^{\frac{5}{8}}}\bigg|^2k~dk=\frac{\hbar}{\alpha}\cdot\frac{8\Gamma^2(\frac{5}{4})\sqrt{2\pi}}{(3w+3)}\cdot\frac{5}{3}\cdot
\frac{1}{a}.
\end{equation}
After plugging the value of $a$ from (\ref{qini}) into
(\ref{Hcrit}), we have:
\begin{equation}
   H_{crit}=\frac{16\Gamma^2(\frac{5}{4})\sqrt{2\pi}}{(3w+3)}\cdot\frac{5}{3}\cdot\frac{1}{\big(\frac{2}{\sqrt{\pi}}
   \Gamma\big(\frac{5+3w}{2+2w}\big)\big)^{w+1}}\frac{E_{tot}}{\hbar}.
\end{equation}
One immediately sees that the value of Hubble parameter at which
the universe `reverses' is proportional to the total energy of
this universe. Therefore, \emph{the more massive the universe, the
higher the value of $H_{crit}$}. Moreover, one may safely assume
that the energy density at the bounce $\rho_{crit}$ is
proportional to the square of the total energy $\sim E_{tot}^2$.

\begin{figure}[t]
\centering
\includegraphics[totalheight=2in]{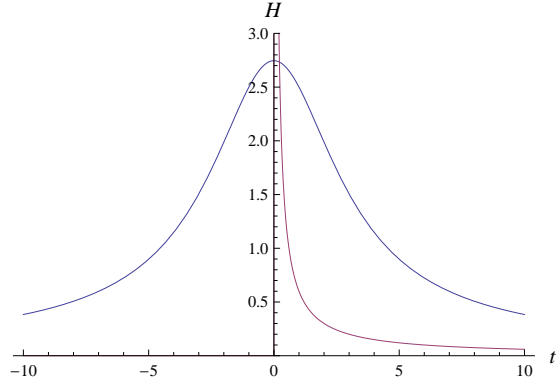}
\caption{The comparison of classical and quantum behavior of the
Hubble parameter for $w=1/3$, $\alpha=\hbar=a=1$. Just after the
Big Bounce the quantum universe is shown to expand faster than the
classical one, $\langle \hat{H}\rangle > H$. Thus, the quantum
universe seems to undergo a kind of inflationary phase.}
\label{fig2}
\end{figure}
\section{Discussion}
\label{d} It is a very interesting conjecture to think that our
phase space with the physical quantities like the Hubble parameter
and the Hamiltonian is in one to one correspondence with the
reduced phase space arising from the identification of Dirac's
observables in the full canonical gauge-symmetry formulation (for
a short comment see appendix \ref{rp}). If so, then our
Hamiltonian corresponds to a true Hamiltonian in the reduced phase
space formulation. This conjecture is investigated in \cite{class}
and a special attention is given to the freedom in fixing an
evolution parameter.

Our analysis shows a way to resolve the singularities of classical
theory. It started with a Lagrangian that was free from gauge
symmetries. The evolution parameter $t$ was treated as a real,
absolute time in which the system evolves. This cosmological clock
does not exist outside the universe and should be connected with
some measurements or change that we observe on cosmological
scales. Nevertheless, no matter which quantization method one uses
for a gauge system with a Hamiltonian itself being a constraint,
one needs an evolution parameter. We argue that the choice of the
evolution parameter in such a way that the singularity occurs at
its finite value is crucial. Thus a popular free scalar field is
not a good choice and we encourage the reader to go through
appendix \ref{B} to see this.

In our approach we managed to express the dynamics of the FRW
universe in terms of a freely-moving non-relativistic particle on
half-line. The classical dynamics ends when the particle hits the
boundary of half-line and the evolution parameter (i.e. the
cosmological time) cannot be extended beyond this moment. However,
we can promote the Hamiltonian to a self-adjoint operator in
quantum theory. Due to the Stone theorem, the self-adjoint
Hamiltonian can be exponentiated to a unitary operator, which
gives the evolution of the system for the unbounded values of the
evolution parameter. Thus, the half-line of the cosmological time
is extended to the whole real domain and the singularity is
resolved.

We have studied the evolution of quantum states $\phi_a$, where
$a\in \mathbb{R}_+$. How much of the Hilbert space is covered by
$\overline{\textrm{Span}}\{\phi_a\}$? We study this question and
others concerning the evolution of unbounded operators
$\hat{\rho}$ and $\hat{H}$ in \cite{quant}.

We observe in the evolution of the Hubble parameter and the volume
of the universe no fixed scale, and in particular Planck scale is
irrelevant. The critical values $v_{crit}$ and $H_{crit}$ depend
on how massive the universe is in the classical phase of its
evolution, which corresponds to the value of $a$. However, we can
calculate the value of $\langle\rho\rangle\langle v^{1+w}\rangle$
at the bounce:
\begin{equation}\label{funsca}
   \alpha H^2_{crit}v^{1+w}_{crit}=\frac{\hbar^2}{\alpha}\sqrt{2\pi}\cdot
   5\cdot(w+1)\bigg(\frac{2\Gamma\big(\frac{5+3w}{2+2w}\big)}{\sqrt{\pi}}\bigg)^{w+1}\Gamma^2\bigg(\frac{5}{4}\bigg),
\end{equation}
which is clearly independent of the parameter $a$, i.e. of a
particular quantum state. Although this quantity is energy
dimensionally it is not the same with the energy of the system.
Notice that the mean value of energy of the system obtained in
(\ref{einv}) depends on a state via the value of $a$. Thus,
(\ref{funsca}) may signal the existence of some fundamental energy
scale not connected with any particular universe. The existence of
such scale may enable to fix the value $\alpha$.

This brings us to the problem of determining the value of
$\alpha$. First, notice that the choice of $\alpha$ is connected
with the choice of time and to this extent it is arbitrary. One
may suspect that the study of the conjectured correspondence
(appendix \ref{rp}) may shed light on this issues and thus we
postpone the investigation to \cite{class}.

Another indeterminacy of the system is connected with the fact the
addition of a total derivative to the Lagrangian in (\ref{action})
leads to the same dynamics. More importantly, there is an infinite
number of ways in which one may define the self-adjoint
Hamiltonian. We have restricted to a single choice, preserving the
Dirichlet condition. But other choices seem to be equally good and
are studied in \cite{quant}.
\begin{acknowledgments}
I want to thank Prof. W. Piechocki for his comments and questions
concerning the ideas presented in this paper.
\end{acknowledgments}
\appendix
\section{Relation between reduced phase space equipped with $\mathbb{H}_{true}$ and reduced Lagrangian $\mathbb{L}_{true}$}
\label{rp}

Suppose one has constructed the reduced phase space equipped with
the induced Poisson bracket and a true Hamiltonian
$\mathbb{H}_{true}$. In principle, one may now construct an action
integral, that is an integral of the Lagrangian
$\mathbb{L}_{true}$ defined as a function of the Dirac
observables:
\begin{equation}
    \mathbb{L}_{true}=P^{J'}\dot{{O}}_{J'}-\mathbb{H}_{true},
\end{equation}
where $P^{J'}$ denote the canonical conjugates to ${O}_{J'}$ (notice that the number of the Dirac observables is always even) and
$\dot{{O}}_{J'}=\{{O}_{J'},\mathbb{H}_{true}\}_{ind}$.

Therefore, one may begin with a reduced form of Lagrangian
$\mathbb{L}_{true}$, which encodes the dynamics of the system in a
gauge fixed in correspondence with $\mathbb{H}_{true}$. This
formulation should also enjoy a freedom in choice of evolution
parameter, hence there should be a class of $\mathbb{L}_{true}$
and a corresponding class of $\mathbb{H}_{true}$.
\section{Massless scalar field case}
\label{B} The metric of FRW universe reads:
\begin{equation}
 ds^2=-N^2(\tau)d\tau^2+a(\tau)^2d\overline{x}^2,
\end{equation}
where $\tau$ is an evolution parameter. The choice of the shift
function $N$ corresponds to the choice of the evolution parameter.
The rest of the notation as well as the meaning of the variable
$l$ used below were explained in section \ref{ct}.
\subsection{Lagrangian formulation}
For the flat FRW universe with massless scalar field the action
reads (see e.g. \cite{wald}):
\begin{equation}\label{amsf}
    S=\frac{3}{16\pi G}\int
    \frac{{\dot{l}}^2l}{N}~dt-\frac{1}{2}\int\frac{l^3{\dot{\phi}}^2}{N}~dt,
\end{equation}
which leads to the Euler-Lagrange equations:
\begin{eqnarray*}
  \frac{\delta S}{\delta l}=0 &\Rightarrow& \frac{3}{16\pi G}\frac{d}{dt}\bigg(\frac{2l\dot{l}}{N}\bigg)-\frac{3}{16\pi G}\frac{\dot{l}^2}{N}+\frac{3}{2}
  \frac{l^2\dot{\phi}^2}{N}=0, \\
  \frac{\delta S}{\delta \phi}=0 &\Rightarrow& \frac{d}{dt}\bigg(\frac{l^3\dot{\phi}}{N}\bigg)=0, \\
  \frac{\delta S}{\delta N}=0 &\Rightarrow& -\frac{1}{N^2}\bigg(-\frac{3}{16\pi
  G}\dot{l}^2l+\frac{1}{2}l^3\dot{\phi}^2\bigg)=0,
\end{eqnarray*}
to which the solutions in the cosmological time ($N=1$) read:
\begin{equation}\label{phi}
    l=\sqrt[3]{\big(8\pi G \Omega_1\big)(t+t_0)},~~~~\phi=\frac{1}{\sqrt{24\pi G}}
    \ln(t+t_0)+\frac{3\Omega_2}{\sqrt{24\pi G}}.
\end{equation}
\subsection{Canonical formulation}
The Legendre transformation followed by Dirac's analysis gives:
\begin{equation}
    \mathbb{H}=N\bigg(\frac{4\pi
    G}{3}\frac{p_l^2}{l}-\frac{1}{2}\frac{p_{\phi}^2}{l^3}\bigg),
\end{equation}
where the phase space is 4D: $(l,p_l,\phi,p_{\phi})$, where
$p_l=\frac{3}{8\pi G}\frac{l\dot{l}}{N}$,
$p_{\phi}=-\frac{l^3\dot{\phi}}{N}$. The Hamiltonian
$\mathbb{H}\approx 0$ is a constraint and $N$ is a non-zero
coefficient corresponding to the choice of gauge. Let us look for
the Dirac observables via fixing the gauge $N=l^3$ and solving:
\begin{equation}\label{dir}
    \{O_i, \mathbb{H}\}=0 \Leftrightarrow
    \bigg[\frac{8\pi G}{3}lp_l\bigg(p_l\frac{\partial}{\partial p_l}-l\frac{\partial}{\partial l}\bigg)+p_{\phi}\frac{\partial}{\partial
    \phi}\bigg]O_i=0.
\end{equation}
We introduce the new 'geometrical' variables $X_{\pm}=\ln
p_l\pm\ln l$ so that the equation (\ref{dir}) reads now:
\begin{equation}
    \bigg[\frac{16\pi G}{3}e^{X_+}\frac{\partial}{\partial
    X_-}+p_{\phi}\frac{\partial}{\partial\phi}\bigg]O_i=0,
\end{equation}
and the solutions are easily found to be:
\begin{equation}
    O_1=X_+,~~~~O_2=p_{\phi},~~~~O_3=\frac{1}{2}X_--\frac{8\pi
    G}{3}\frac{e^{X_+}}{p_{\phi}}\phi ,
\end{equation}
which are not independent on the constraint surface due to the identity $\mathbb{H}=\frac{4\pi G}{3}e^{2O_1}-\frac{1}{2}O_2^2$ and for which the
Poisson bracket is now modified as follows $dl\wedge dp_l=\frac{1}{2}e^{X_+}dX_+\wedge dX_-$. So the complete set of independent elementary
Dirac observables consists of two elements, say $O_1$ and $O_3$, satisfying the algebra:
\begin{equation}
    \{O_1,O_3\}=e^{-O_1}.
\end{equation}
We redefine $\Omega_1=e^{O_1}$ and
$\Omega_2=O_3-\frac{1}{3}\ln\frac{\sqrt{\Omega_1}}{8\pi G}$, so
that
\begin{equation}
    \{\Omega_1,\Omega_2\}=1,
\end{equation}
and we restrict the analysis to the `expanding universe' solutions
$\Omega_1=lp_l>0$ and on the constraint surface we have
$O_3=\frac{1}{2}X_-+\sqrt{\frac{8\pi G}{3}}\phi$. From (\ref{phi})
we have that:
\begin{equation}\label{length}
    l_{\phi}=\sqrt[3]{8\pi G \Omega_1e^{-3\Omega_2}}e^{\sqrt{\frac{8\pi
    G}{3}}\phi}.
\end{equation}
This quantity is now the Dirac observable for a fixed value of
$\phi$, since it does not depend on the value of $t$ that one
ascribes to the moment of measurement. The evolution of length is
defined as the flow through the parameter $\phi$, which
parameterizes the family of Dirac's observables $l_{\phi}$.

\subsection{Classical evolution}

Let us identify the generator of the change in $\phi$:
\begin{equation}\label{h1}
\{ l_{\phi},
\mathbb{H}_{\phi}\}=\partial_{\phi}l_{\phi}\Leftrightarrow
\frac{1}{3\Omega_1}\frac{\partial \mathbb{H}_{\phi}}{\partial
\Omega_2}+\frac{\partial \mathbb{H}_{\phi}}{\partial
\Omega_1}=\sqrt{\frac{8\pi G}{3}}\Leftrightarrow
\mathbb{H}_{\phi}=\sqrt{\frac{8\pi
G}{3}}\Omega_1+f(\frac{1}{3}\ln\Omega_1-\Omega_2),
\end{equation}
where $f$ is arbitrary. How to get rid of this indeterminacy? Note that to parameterize the reduced phase space one needs two independent
Dirac's observables. The first choice was $l_{\phi}$ and what is to be the second? Let us obtain from (\ref{phi}) the Hubble parameter:
\begin{equation}\label{hublle}
H_{\phi}=\frac{1}{3}e^{3\Omega_2}e^{-3\sqrt{\frac{8\pi
G}{3}}\phi},
\end{equation}
so that
\begin{equation}\label{h2}
\{ H_{\phi},
\mathbb{H}_{\phi}\}=\partial_{\phi}H_{\phi}\Leftrightarrow
\frac{\partial \mathbb{H}_{\phi}}{\partial
\Omega_1}=\sqrt{\frac{8\pi G}{3}}\Leftrightarrow
\mathbb{H}_{\phi}=\sqrt{\frac{8\pi G}{3}}\Omega_1+g(\Omega_2),
\end{equation}
where $g$ is arbitrary. Now, comparing (\ref{h1}) and (\ref{h2}),
removes the indeterminacy and the true Hamiltonian is
$\mathbb{H}_{\phi}:=\sqrt{\frac{8\pi G}{3}}\Omega_1$. Thus,
instead of considering the whole family of $l_{\phi}$ and
$H_{\phi}$, which of course produce redundancy in parameterizing
the reduced phase space, one fixes the value of $\phi$ and
includes the $\mathbb{H}_{\phi}$ as a dynamics generator. One only
needs to include that:
\begin{equation}
\{H_{\phi},l_{\phi}\}=-\frac{8\pi G}{3}l_{\phi}^{-2},~~\{
l_{\phi}, \mathbb{H}_{\phi}\}=\sqrt{\frac{8\pi G}{3}}l_{\phi},~~\{
H_{\phi}, \mathbb{H}_{\phi}\}=-3\sqrt{\frac{8\pi G}{3}}H_{\phi},
\end{equation}
where $\mathbb{H}_{\phi}=\sqrt{\frac{3}{8\pi G}}H_{\phi}l_{\phi}^3$. Now, let us identify $p_{\phi}$, the conjugate of $l_{\phi}$:
\begin{equation}
    \{l_{\phi}, p_{\phi}\}=1\Leftrightarrow p_{\phi}=(8\pi G)^{-\frac{1}{3}}e^{-\sqrt{\frac{8\pi G}{3}}\phi}\Omega_1^{\frac{2}{3}}e^{\Omega_2}
    +f(\Omega_1e^{-3\Omega_2})=\frac{3}{8\pi
    G}l_{\phi}^2H_{\phi}+f(l_{\phi}),
\end{equation}
where $f$ is arbitrary but we put $f\equiv 0$.
\subsection{Quantum evolution}
Now, let us collect the information on the system in the form convenient for quantization:
\begin{equation}
 \{l, p_{\phi} \}=1,~~~~\mathbb{H}=\sqrt{\frac{8\pi
 G}{3}}lp_{\phi},~~~~H=\frac{8\pi
 G}{3}\frac{p_{\phi}}{l^2},~~~~\rho=\frac{8\pi
 G}{3}\frac{p_{\phi}^2}{l^4},~~~~v=l^3,
\end{equation}
where $l>0$, we have dropped the index ${\phi}$ for convenience and thus $l$, $\mathbb{H}$, $H$ and $\rho$ are the length, the Hamiltonian, the
Hubble parameter and the energy density, respectively.

We do the following canonical transformation:
\begin{equation}
   l\mapsto x= \frac{1}{3}l^{3},~~~~p_{\phi}\mapsto
   p=\frac{p_{\phi}}{l^2},
\end{equation}
where the new canonical pair $\{x,p\}=1$ is also defined on
$\mathbb{R}_{+}\times \mathbb{R}$ and
\begin{equation}
\mathbb{H}=\sqrt{24\pi
 G}xp,~~~~H=\frac{8\pi
 G}{3}p,~~~~\rho=\frac{8\pi
 G}{3}p,~~~~v=3x.
\end{equation}
Now, the Schr\"odinger quantization, $p\mapsto \hat{p}=:-i\frac{d}{dx}$ and $x\mapsto \hat{x}=x$, renders the following formally self-adjoint
operators:
\begin{equation}
    \hat{\mathbb{H}}=\sqrt{6\pi
 G}\bigg(-ix\frac{d}{dx}-i\frac{d}{dx}x\bigg),~\hat{H}=\frac{8\pi
 G}{3}\bigg(-i\frac{d}{dx}\bigg),~\hat{\rho}=\frac{3}{8\pi
 G}\hat{H}^2,~\hat{v}=3x.
\end{equation}
Let us consider the following unitary mapping:
\begin{equation}
R_+\ni x,~~\Psi(x)\mapsto\Phi(y)=\sqrt{2}e^y\Psi(e^{2y}),~~y\in R,
\end{equation}
and we find that the Hamiltonian operator in this new Hilbert space reads:
\begin{equation}
     \hat{\mathbb{H}}=\sqrt{6\pi
 G}\big(-i\frac{d}{dy}\big).
\end{equation}
Now we should ensure that the Hamiltonian is a positive operator,
but we will ignore it for it will not alter the fate of the
singularity. Now, we can obtain the evolution of wave-functions:
\begin{equation}
e^{i\mathbb{H}\phi}\Phi(y)=\Phi(y+\phi\sqrt{6\pi
 G})=\sqrt{2}e^ye^{\phi\sqrt{6\pi
 G}}\Psi(e^{2y}e^{2\phi\sqrt{6\pi
 G}}),
\end{equation}
hence the inverse mapping back to the start Hilbert space gives:
\begin{equation}
\sqrt{2}e^ye^{\phi\sqrt{6\pi
 G}}\Psi(e^{2y}e^{2\phi\sqrt{6\pi
 G}})\mapsto e^{\phi\sqrt{6\pi
 G}}\Psi(e^{2\phi\sqrt{6\pi
 G}}x).
\end{equation}
Now, we are ready to study the evolution of the expectation value of $\hat{\rho}$:
\begin{equation}
\langle e^{i\mathbb{H}\phi}\Psi(x)|\hat{\rho}~e^{i\mathbb{H}\phi}\Psi(x)\rangle=\frac{8\pi G}{3}e^{2\sqrt{24\pi
 G}\phi}\int \overline{\Psi}(x)-\frac{d^2}{dx^2}\Psi(x)~dx,
\end{equation}
and the expectation value of $\hat{v}$:
\begin{equation}
\langle e^{i\mathbb{H}\phi}\Psi(x)|\hat{v}~e^{i\mathbb{H}\phi}\Psi(x)\rangle=3e^{-\sqrt{24\pi
 G}\phi}\int \overline{\Psi}(x)x\Psi(x)~dx.
\end{equation}
The classical behavior is reproduced in the quantum theory, which
includes  the unbounded growth of energy density and the vanishing
size of the universe.

One may think that this shows that the singularity cannot be
resolved in the above system. In fact, however, the singularity
problem was not addressed at all due to the peculiar choice of the
evolution parameter $\phi$. In order to study the structure of the
singularity one has to ensure that the classical singularity
occurs at finite value of an evolution parameter.

\end{document}